\documentclass[aps,prl,twocolumn,showpacs,groupedaddress]{revtex4}
\usepackage{epsfig}
\usepackage{dcolumn}   
\usepackage{bm}        
\usepackage{amssymb}

\begin{document}
\newcommand{\mybibitem}{\item}

\title{Constraints on AdS/CFT Gravity Dual Models of Heavy Ion Collisions}

\author{Jorge Noronha$^1$, Miklos Gyulassy$^{1}$, and Giorgio Torrieri$^{2}$}
\affiliation{$^1$Department of
Physics, Columbia University, 538 West 120$^{th}$ Street, New York,
NY 10027, USA\\
$^2$Frankfurt Institute for Advanced Studies, Frankfurt am Main, Germany}

\begin{abstract}
We show that the five-fold constraints due to (1) the observed nuclear
modification of heavy quark jets measured via non-photonic electrons
$R_{AA}^e(p_T \sim 6\,{\rm GeV})$ in central Au+Au collisions at 200
AGeV, (2) the ``perfect fluid'' elliptic transverse flow of low
transverse momenta pions, $v_2(p_T\sim 1\;{\rm GeV})$ reported for
noncentral collisions, (3) the global pion rapidity density
$dN_\pi/dy$, (4) the lattice QCD entropy density deficiency,
$S/S_{SB}$, of strongly coupled Quark-Gluon Plasmas (sQGP),  and (5)
a causal requirement are analytically correlated in a class of gauge/string
gravity dual models of sQGP dynamics. Current RHIC/BNL
and lattice QCD data are found to be remarkably
compatible with these models if the
t'Hooft and Gauss-Bonnet coupling parameters lie in the range
$\lambda\approx 10-25$ and $0< \lambda_{GB}< 0.09$. In addition, the
observed five-fold correlation appears to favor color glass
condensate over Glauber initial conditions within current
systematic errors.
\end{abstract}


\date{\today}
\maketitle



The combined observations of the quenching of hard (high transverse
momentum or high quark mass jets) processes and the nearly ``perfect
fluid'' elliptic flow of soft (low momentum transfer) hadrons produced
in Au+Au collisions at $\sqrt{s}=200$ AGeV at the Relativistic Heavy
Ion Collider (RHIC) \cite{RHICwhitepapers} have been interpreted as
providing evidence for the formation of a new form of strongly
interacting quark-gluon plasma (sQGP) \cite{Gyulassy:2004zy}. Furthermore, bulk multiplicity (entropy)
production systematics have been interpreted as evidence for
gluon saturation of the initial conditions as predicted by the Color
Glass Condensate (CGC) model
\cite{Kharzeev:2001yq,CGCinitialcondition}.
However, it has
been a challenge to find a single consistent theoretical framework that
can explain simultaneously both soft and hard phenomena.  These
phenomena
include 1) the
nuclear modification of high transverse momenta, $p_T > 5$ GeV, jet
observables, 2) the bulk collective flow observables ($p_T < 1$ GeV),
and 3) the sQGP thermodynamic equation of state entropy deficiency
relative to the ideal Stefan-Boltzmann limit as
predicted by nonperturbative Lattice QCD (LQCD). Attempts to
explain these based on
weak coupling (perturbative QCD) parton transport approaches
\cite{Danielewicz:1984ww,coesterpaper,majumder,fochler}, especially the surprising small viscosity needed to fit the elliptic
flow data, require large coupling
extrapolations $\alpha_s=g_{YM}^2/4\pi \rightarrow 0.5-0.6$.  At such large gauge couplings, on the other hand, the t'Hooft
parameter $\lambda=g_{YM}^2 N_c \gg 1$ may already be large enough to
validate string theory inspired AdS/CFT low energy approximations
\cite{maldacena,Gubser:1998nz}. The great theoretical advantage and
appeal of gravity/gauge dual models \cite{maldacena}-\cite{iranian} is
that they have led (for the first time) to {\em analytic} connections
between a wide variety of thermodynamic and nonequilibrium dynamic
variables at strong coupling that were not yet realized with traditional gauge theory
techniques.

In this Letter, we focus on predicted analytic connections between
three fundamental properties of the sQGP: (1)
its equation of state (entropy=S), (2) its long wavelength transport coefficients
(viscosity=$\eta$) , and (3) coupling between
long wavelength near-equilibrium ``soft medium'' properties and
short-wavelength non-equilibrated ``hard probes''
(energy loss per unit length, $dE/dx)$.
We show that these
three properties together with global entropy and causality restrictions
can provide valuable phenomenological constraints
of higher dimensional gravity dual models of sQGP in
heavy ion collisions.
We consider here the constraints imposed by current RHIC and LQCD data on a class of gravity dual models
that include quadratic as well as quartic curvature corrections to the classical Einstein-Hilbert action for the effective 5 dimensional
dual gravity action. Therefore, we implement and extend the suggestion made in \cite{Buchel:2008vz} to
include perturbatively both the lowest order $O(\lambda_{GB}\sim 1/N_c)$
Gauss-Bonnet $\mathcal{R}^2$ and $O(1/\lambda^{3/2})$ $\mathcal{R}^4$
curvature corrections to the three properties above.
\begin{figure}[h]
\epsfig{file=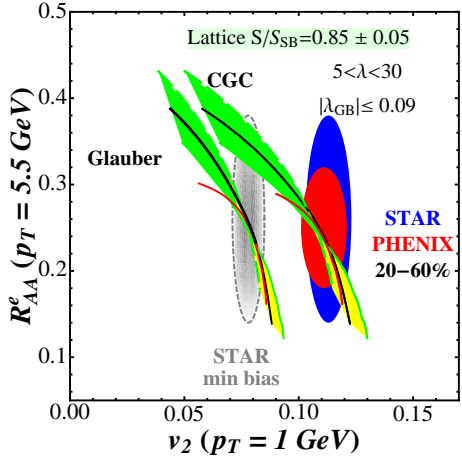,width=2.5in,clip=}
\caption{\label{Raaxv2plot} (Color online) The gravity dual model
correlation of the nuclear modification factor
  , $R_{AA}^e$, of hard ($p_T=5.5$ GeV, 0-10\% centrality) non-photonic
  electrons with the elliptic transverse flow moment,  $v_2$, of soft ($p_T=1$ GeV,
  20-60\% centrality) pions assuming Glauber (left)
  and CGC (right) initial participant distributions and  $dN_{\pi}/dy=1000$
are compared. The green
  region corresponds to the range of entropy function deficiencies,
  $S/S_{SB}$, consistent with infinite volume extrapolations of lattice QCD data
  \cite{latticedata}.  The red
  (blue) ellipse indicates PHENIX (STAR) 0-10\% centrality $R_{AA}^e$
  and 20-60\%$v_2$ centrality Au+Au 200 AGeV data
  from RHIC  \cite{Adler:2005xv}-\cite{v2data}. The dashed blue
  curves show fixed $S/S_{SB}=0.8$ while for the solid black
  and dotted-dashed curves $S/S_{SB}=0.85$ and $S/S_{SB}=0.9$,
  respectively in the range $\lambda=5, \ldots, 30$. The causality bound curve
  with $\lambda_{GB}=0.09$ is indicated by solid red. The gray ellipse
  shows preliminary $v_2$ for minimum bias data with estimated non-flow effects
  subtracted \protect\cite{poszkanzer}.
}
\end{figure}

Our analysis is based on the following remarkably simple
algebraic expressions relating these three fundamental sQGP  properties:
\begin{eqnarray}
& S/S_{SB} & = \frac{3}{4}\left(1+\lambda_{GB}+\frac{15}{8}\frac{\zeta(3)}{\lambda^{3/2}}\right),
\label{sbargeneral}\\
& \eta/s & = \frac{1}{4\pi}\left(1-4\lambda_{GB}+15\frac{\zeta(3)}{\lambda^{3/2}}\right),
\label{etasgeneral}\\
& \tau_Q^{-1} 
&
= \mu_{Q} \left(1+\frac{3}{2}\lambda_{GB} + \frac{15}{16}\frac{\zeta(3)}{\lambda^{3/2}} \right)
\label{dpdtgeneral}
\end{eqnarray}
where $s=S/V$ is the entropy density.
The heavy quark jet relaxation rate, $1/\tau_Q$, is
controlled by $\mu_{Q}=\sqrt{\lambda}\,\pi
T^2/2M_{Q}$ for a heavy quark with mass $M_Q$ in a plasma of
temperature $T$ .
The relaxation time is related to the energy loss per unit length
through $\tau_Q(\lambda,\lambda_{GB})= -1/(d\log
p/dt)=-1/(d\log E/dx)$, where $p=M_{Q}\gamma v$ and $v=p/E$.

We note that the $\mathcal{R}^4$ correction $O(1/\lambda^{3/2})$ to
the heavy quark jet energy loss is a new result \cite{noronhanext} reported in
this Letter and it is needed for a consistent additive perturbative
application of this class of AdS/CFT models to heavy ion reactions. We
consider here only heavy quark observables because gravity dual string drag
models \cite{drag} apply only for heavy $M_Q> \sqrt{\lambda} T$ jets.
The heavy quark jet drag is then modeled as a
trailing string moving in a black brane background
according to the classical Nambu-Goto action (with dilaton neglected)
$ \mathcal{A}_{NG}=-\frac{1}{2\pi \alpha'}\int d^2\sigma \sqrt{-g}
$
where $g=
  \det\,g_{ab}=G_{\mu\nu}\partial_{a}X^{\mu}\partial_{b}X^{\nu}$ is
the induced worldsheet metric, $\sigma^{a}=(\tau,\sigma)$ are the
internal worldsheet coordinates, $G_{\mu\nu}(X)$ is the background metric,
 and $X^{\mu}=X^{\mu}(\tau,\sigma)$ is
the embedding of the string in spacetime. The trailing string
ansatz (where $\tau=t,\,\sigma=u$ and
$X^{\mu}(t,u)=(t,x_{0}+vt+\xi(u),0,0,u)$) describes the asymptotic
behavior of a string attached to a moving heavy quark (the string
endpoint) with velocity $v$ in the $x$ direction and located at a
fixed AdS radial coordinate $u_m\gg u_h$ \cite{drag}.
The black brane horizon coordinate $u_h \propto T \alpha'$ is
is determined by
$G_{00}(u_h)=0$. Using the ansatz above and the string's classical
equations of motion, one can show that the drag force $dp/dt= - C
v/(2\pi \alpha')$, where $C$ is a constant determined by the
negativity condition
that
\begin{equation}
\ g(u) = G_{uu}\left(G_{00}+v^2 G_{xx}\right)\left(1+\frac{C^2 v^2}{G_{00}G_{xx}}\right)^{-1}
<0
\label{inducedmetric}
\end{equation}
for $u_h \leq u \leq u_m$. However, both the numerator and
denominator in Eq.\ (\ref{inducedmetric}) change their sign
simultaneously at a certain $u^*$ \cite{drag} given by the root of
the  equation
$
\ G_{00}(u^*) +v^2 G_{xx}(u^*)=0
$. This fixes $C=G_{xx}(u^*)$ and 
$dp/dt=-v
\,G_{xx}(u^*)/(2\pi \alpha')$. Neglecting higher-order derivative
corrections in $\mathcal{N}=4$ SYM one finds $u^*=u_h \sqrt{\gamma}$,
where $\gamma=1/\sqrt{1-v^2}$. The condition that $u^*\leq u_m$
leads to a maximum ``speed limit'' for the heavy quark jet to be
consistent
with this trailing string ansatz given by $\gamma_{\rm
  max}\leq u_m^2/u_h^2$ \cite{speedlimit}.

Using the metric derived in \cite{Gubser:1998nz} to $\mathcal{O}(\alpha\,'^{3})$, one can compute the effects of quartic corrections on the drag force \cite{poritz} and determine $u^*$ perturbatively to $\mathcal{O}(\lambda^{-3/2})$ as \cite{comment}
$ 
 u^* =u_h \sqrt{\gamma} \left[1+ \frac{15}{32}\frac{\zeta(3)\, v^2}{\lambda^{3/2}}\left(5+\frac{5}{\gamma^2}-\frac{3}{\gamma^4}\right)\right]
$
and the drag force
\begin{equation}
\frac{dp}{dt} = -\sqrt{\lambda}\,T^2\frac{\pi}{2} v\gamma \left[1+\frac{15}{16}\frac{\zeta(3)}{\lambda^{3/2}}\left(1-\frac{197}{24\gamma^4}+\frac{67}{24\gamma^6}\right)\right] .
\label{newdpdt}
\end{equation}
The heavy quark mass at $T=0$ is $M_Q=u_m/(2\pi \alpha')$ and, to leading order in $1/\lambda$, $u_m^2/u_h^2\simeq \frac{4M_Q^2}{\lambda T^2}$. Thus, the corrected $u^*$ displayed above defines a new speed limit $\gamma_m \simeq \frac{4M_Q^2}{\lambda T^2}\left[1-\frac{5}{16}\left(\frac{4\pi\eta}{s}-1\right)\right]$, after neglecting terms of $\mathcal{O}(1/\gamma,1/N_c)$. Note that $\gamma_m$ and $dp/dt$ decrease with increasing $\eta/s$.

For our applications,  we consider the range $\lambda\sim 5-30$ and $|\lambda_{GB}|<0.1$
with fixed $N_c=3$. In this parameter range the $1/\lambda^{3/2}$ and
$\lambda_{GB}\sim 1/N_c$ corrections are comparable.
We neglect known (but formally) higher order
terms \cite{Buchel1} $\mathcal{O}(\sqrt{\lambda}/N_c^2)$
in this first attempt to test predicted dynamical
correlations between hard and soft phenomena in high energy A+A collisions.
\begin{figure}[t]
\hspace{-.07in}\epsfig{file=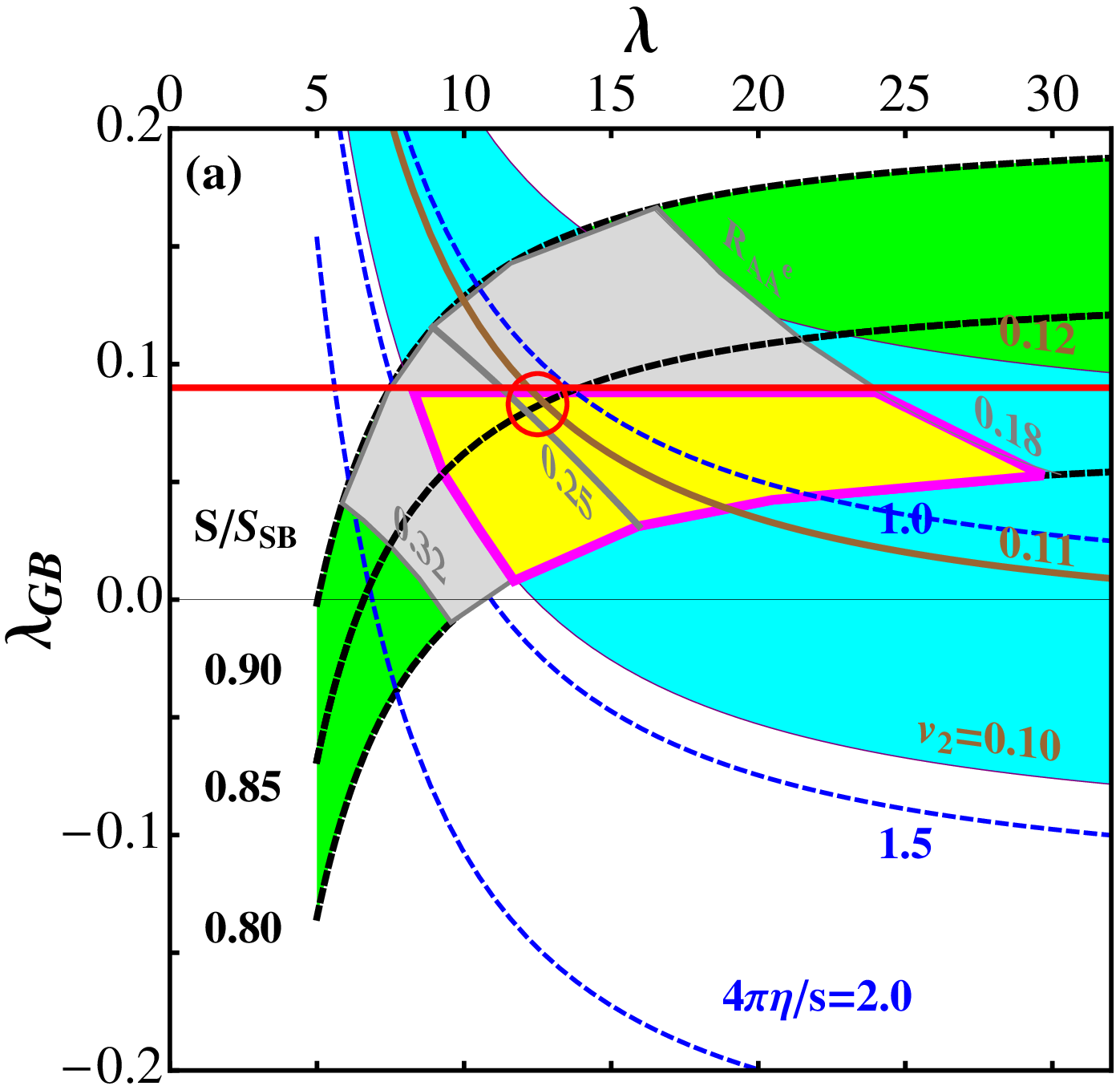,width=2.58in,height=1.8in,clip=}
\vspace{-0.22in}
\hspace{.08in}
\epsfig{file=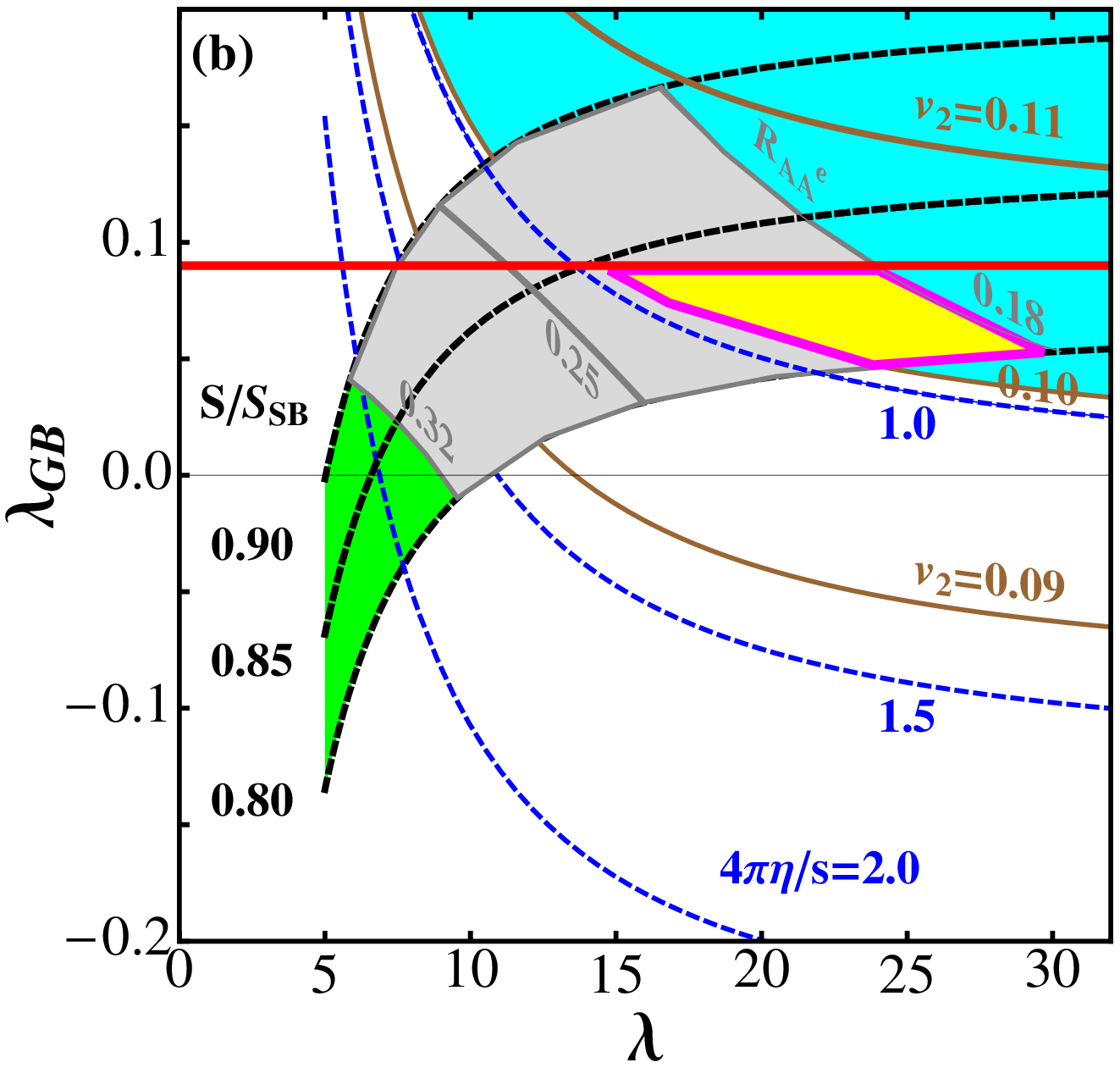,width=2.75in,height=1.8in,clip=}
\caption{\label{lambdalambgbplot} (Color online) Five fold phenomenological
constraints in the t'Hooft and Gauss-Bonnet parameter space
$(\lambda,\lambda_{GB})$. (1) The green region with black dashed contours
is from lattice QCD constraints on $0.8<S/S_{SB}<0.9$.
(2) The cyan region is determined from noncentral elliptic
flow $v_2(p_T=1,20-60\%)=0.11\pm0.01$
\protect\cite{v2data}. Blue dashed contours correspond to fixed
$4\pi\eta/s=1,1.5,2$.
The inversion $\lambda_{GB}(\lambda,v_2(\eta/s))$ is
based on minimum bias viscous hydro results of \protect\cite{Luzum:2008cw}
assuming CGC initial eccentricities scaled by a factor 1.1 in panel (a)
and unscaled 1.0 in panel (b). The entropy is constrained
by the (3) $dN_\pi/dy=1000$ pion rapidity density in central collisions.
(4) The gray region and contours
are determined from central $R_{AA}^e(p_T=5.5\,{\rm GeV})=0.25\pm0.07$
data \protect\cite{Adler:2005xv}.
(5) The horizontal red line constraint
is the causality upper bound for
 $\lambda_{GB}^{max}=0.09$ \protect\cite{brigantecausal,Hofman:2008ar}.
The yellow trapezoidal
region with the purple boundary is the intersection
of the five fold constraint bands.
 Note the red circled five fold
conjunction area in panel (a)
$(\lambda\approx 13,\lambda_{GB}\approx 0.08)$ that is absent in the
unscaled
panel (b).}
\end{figure}

The small Gauss-Bonnet parameter $\lambda_{GB}=(c-a)/4c$ is related to the
central charges $c$ and $a$ (related to the conformal anomaly in curved spacetime) of the
dual CFT as noted in Eq.\ (2.14) of Ref.\ \cite{Buchel:2008vz}. Varying
$\lambda_{GB}$ provides a parametric way to explore deformations of the original
${\cal N}=4\; SU(N_c)$ SYM theory. Interest in Gauss-Bonnet deformations were heightened when Kats and Petrov \cite{Kats:2007mq}
argued that for ${\cal N}=2\; Sp(N_c)$, $\lambda_{GB}= 1/8N_c$, the KSS viscosity bound on $\eta/s\ge 1/4\pi$ was violated
by 17\% for $N_c=3$. As further shown in \cite{Buchel:2008vz}, a large class of other effective CFTs are now known to lead to similar $\lambda_{GB}\propto 1/N_c$ effects. However, the analysis of Refs.\ \cite{Brigante:2007nu}-\cite{Hofman:2008ar}
revealed that $\lambda_{GB}$ deformations are limited by requirements of causality and positive energy flow to a narrow range $-7/36<\lambda_{GB}<9/100$.

In order to convert $\tau_Q$ into the observed nuclear
modification of single non-photonic electrons, $R_{AA}^e(p_T=5.5 \;
{\rm GeV})$, from quenched heavy quark jets,
we follow and extend \cite{Horowitz:2007su} by using here
the generalized drag force
in Eq.\ (\ref{dpdtgeneral}) to compute the path length, $L$,
dependent heavy quark
fractional energy loss $\epsilon(L)$.
The heavy quark jet nuclear modification factor is then
$R_{AA}=\langle(1-\epsilon)^{n_Q}\rangle_L$, where $n_{Q}(p_T)$
is the flavor dependent spectral index
$n_{Q}+1=-\frac{d}{d\,\ln
  p_T}\ln\left(\frac{d\sigma_{Q}}{dydp_T}\right)$ obtained from FONLL
production cross sections \cite{cacciari} as used in
\cite{Horowitz:2007su}.
The path length average of
the nuclear modification at impact parameter $b$ is computed using a
Woods-Saxon nuclear density profile with Glauber profiles
$T_A(\vec{x}_{\perp})$ with $\sigma_{NN}=42$ mb. For 0-10\% centrality
triggered data both Glauber and CGC geometries lead to
similar numerical results \cite{Drescher:2006pi}. The distribution of initial hard jet
production points at a given $\vec{x}_{\perp}$ and azimuthal direction
$\phi$ is taken to be proportional to the binary parton collision
density, $T_{AA}(\vec{x}_{\perp},b)$.
We assume a longitudinally expanding local (participant) parton density
$\rho(\vec{x}_{\perp},b)=\chi
\rho_{part}(\vec{x}_{\perp},b)/\tau$, where $\chi\equiv (dN_{\pi}/dy)/N_{\rm
  part}$ and $\rho_{part}$ is the Glauber participant nucleon profile density.
However, we evaluate (\ref{dpdtgeneral}) with
a reduced temperature $T_{CFT}=0.74 (S/S_{SB})^{1/3}T_{QCD}$
to take into account the smaller number of degrees of freedom
in a strongly-coupled QCD plasma, which is similar to the prescription given in \cite{Gubser:2006qh}.
We compute the heavy quark modification factor
$R_{AA}^Q$ via
\begin{eqnarray}
\ R_{AA}^{Q}(p_T,b)&=&\int_{0}^{2\pi}d\phi\int
d^2\vec{x}_{\perp}\frac{T_{AA}(\vec{x}_{\perp},b)}{2\pi\, N_{\rm
    bin}(b)} \nonumber \\ &\;& \hspace{-1 in} \times
\exp\left[-n_{Q}(p_T)\int_{\tau_0}^{\tau_f}
  \frac{d\tau}{\tau_c(\vec{x}_{\perp}+\tau \hat{e}(\phi),\phi)}\right]
\label{Raaq}
\end{eqnarray}
where $N_{\rm bin}$ is the number of binary collisions.
Here, $\tau_0=1$ fm/c is the assumed plasma equilibration time and $\tau_f$
is determined from $T(\vec{\ell},\tau_f)=T_f=140$
MeV, i.e, the time at which the local temperature falls below a freeze-out
temperature taken from \cite{Luzum:2008cw}.
Systematic errors associated with the freeze-out
condition will be discussed elsewhere.

In order to compute $v_2(1\; {\rm GeV})$ for the $\mathcal{C}$(20-60\%) centrality
class, we employ a linear fit to the numerical viscous hydrodynamic results of Luzum and Romatschke \cite{Luzum:2008cw}.
The dependence of $v_2$ on viscosity for both Glauber
\cite{glauber} and CGC \cite{CGCinitialcondition} initial transverse profiles can be well fit with
\begin{equation}
v_2(p_T,\eta/s,\mathcal{C})= a(p_T)\, \epsilon_2(\mathcal{C}) \,( 1- b\,\eta/s)
\end{equation}
where $\epsilon_2(\mathcal{C})=\langle y^2-x^2\rangle_{\mathcal{C}} /\langle x^2+y^2\rangle_\mathcal{C}$
is the average initial elliptic geometric eccentricity for the centrality class $\mathcal{C}$. To rescale the minimum bias viscous hydro results of Ref.\ \cite{Luzum:2008cw} to the considered 20-60\% centrality class
we use the factor $\epsilon^{Glaub}_2(20-60\%)/\epsilon^{Glaub}_2(0-92\%)=0.317/0.281=1.128 $
from Ref.\ \cite{Adcox:2002ms}. Our fit to the rescaled numerical results of \cite{Luzum:2008cw}
give $b\approx 2.5$ and $a(p_T=1)\epsilon_2(20-60\%)\approx 0.14 \;(0.098)$
for CGC (Glauber) initial conditions. We consider this 20-60\% centrality class
because, as shown in Fig.\ 23 of \cite{Afanasiev:2009wq}, there is good agreement at $p_T\sim 1$ GeV
between STAR $v_2(4)$ and PHENIX $v_2(BBC)$ data
and non-flow effects \cite{poszkanzer} that complicate the interpretation
of minimum bias data in \cite{Luzum:2008cw} are reduced.

We find (see Figs.\ \ref{Raaxv2plot} and \ref{lambdalambgbplot}) that
a (reasonable) combination of model parameters ($\lambda$ and
$\lambda_{GB}$) can account for the correlation between the reported
$R_{AA}^{e}$ \cite{Adler:2005xv,starelectronRaa} and $v_2$
\cite{v2data} taking into account the LQCD constraint on the equations
of state deficiency $S/S_{SB}$ \cite{latticedata} (in the range $T\sim
2-4 T_c$). Our results suggest  that within current
systematic errors  a small positive quadratic curvature
correction with  $0<\lambda_{GB}<0.1$ is preferred phenomenologically.
Moreover, CGC initial eccentricities are favored over Glauber eccentricities  for the
centrality class considered. However, as emphasized by Fig.\ 2b, the
results are rather sensitive to the
$\lambda_{GB}(\lambda,v_2(\eta/s))$ inversion. We find that a  10\% reduction
of the viscous hydrodynamic $v_2(\eta/s)$ (as for minimum bias
centrality) virtually eliminates the yellow overlap region and could
falsify the AdS/CFT description based on Eqs.\ (1-3). Great care is
called for at this time to avoid premature conclusions.
Systematic studies on viscous hydrodynamic dependence on centrality cuts
and improved experimental control over non-flow corrections will
be needed before definitive conclusions could be reached. The
good news demonstrated by comparing Fig. 2a and 2b is that if
improved theoretical and experimental control (better than 10\%) over
elliptic flow systematics and the nuclear modification of heavy quark
jet observables can be reached, then rather strong experimental constraints on
the AdS/CFT gravity dual model parameters  could be achieved.
We close by emphasizing \cite{Horowitz:2007su}
that future comparison of the nuclear modification of {\em identified} bottom and charm quark jets at RHIC and LHC
combined with the fivefold (hard/soft) constraints
considered in this Letter will provide especially stringent
tests of AdS/CFT gravity dual phenomenology applied to
 high energy heavy ion reactions.

We thank A.\ Dumitru, S.\ Gubser,  W.\ Horowitz, A. Poszkanzer, B.~Cole, and W.~Zajc for useful comments. J.N. and M.G. acknowledge support from US-DOE Nuclear Science Grant No. DE-FG02-93ER40764. G.T. acknowledges support from the Helmholtz International
Center for FAIR within the framework of the LOEWE program
(Landesoffensive zur Entwicklung Wissenschaftlich-\"Okonomischer
Exzellenz) launched by the State of Hesse.


\begin{thebibliography}{99}
\bibitem{comment} The drag force in this case had been computed numerically in Ref.\ \cite{poritz} but the analytical expression in Eq.\ (\ref{newdpdt}) is, as far as we know, a new result.
\end{thebibliography}
\end{document}